\documentclass[Journal]{IEEEtran}
\usepackage{graphicx,epstopdf}
\usepackage{amssymb,amsmath}
\usepackage{commath}
\usepackage{setspace}
\usepackage{acronym}
\usepackage{mathrsfs}
\usepackage{ifthen}
\usepackage{textcomp}
\usepackage{float}
\usepackage{subfigure}
\usepackage{color}
\usepackage{cite}
\usepackage{makecell}

\IEEEoverridecommandlockouts

\begin{document}

\title{Demo: A Reinforcement Learning-based Flexible Duplex System for B5G with Sub-6 GHz}


\author{ 
	Soo-Min~Kim,~\IEEEmembership{Student~Member,~IEEE,}
	Han~Cha,~\IEEEmembership{Student~Member,~IEEE,}\\
	Seong-Lyun~Kim,~\IEEEmembership{Senior~Member,~IEEE,}
	and~Chan-Byoung Chae,~\IEEEmembership{Senior~Member,~IEEE}\\
	Yonsei University, Korea,
	email: cbchae@yonsei.ac.kr
}

 \maketitle

\begin{abstract}
In this paper, we propose a reinforcement learning-based flexible duplex system for B5G with Sub-6 GHz. This system combines full-duplex radios and dynamic spectrum access to maximize the spectral efficiency. We verify this method's feasibility by implementing an FPGA-based real-time testbed. In addition, we compare the proposed algorithm with the result derived from the numerical analysis through system-level evaluations.
\end{abstract}

\section{Motivation and Preliminaries}
The scarcity of the licensed spectrum is also a never-ending subject in 5G and beyond 5G (B5G) networks. These phenomena have fueled a growing interest in solutions that tackle spectral efficiencies such as dynamic spectrum access (DSA) and full-duplex radios~\cite{Liao2015LAT,Sachin2013Full}. 
DSA exploits the spectrum holes to maximize spectral efficiency, while full-duplex radios transmit and receive the signals simultaneously at the same frequency to maximize it. 
Utilizing both technologies has also been actively studied recently, which is called a flexible duplex system~\cite{dyspandemo,opmapfd}. 
However, it still faces a few limitations. The flexible duplex system decides whether to transmit or not based on the opportunity (OP) map~\cite{op}, aided by the stochastic geometry. To this end, we define the proper access threshold value that determines the boundary of the transmission depending on the environment and situation. Determining this value in real-time is a difficult task. 

To handle these obstacles, we share a new solution in this paper called ``reinforcement learning-based flexible duplex system” in a sensor-aided mode.
This research's main contribution lies in providing the characteristics of the flexible duplex system that utilizes a reinforcement learning algorithm in a real-time prototype.
More analyses and algorithms will be described in an extended draft.

\section{Testbed Design}




 \begin{figure}[t]
	\centerline{\resizebox{0.95\columnwidth}{!}{\includegraphics{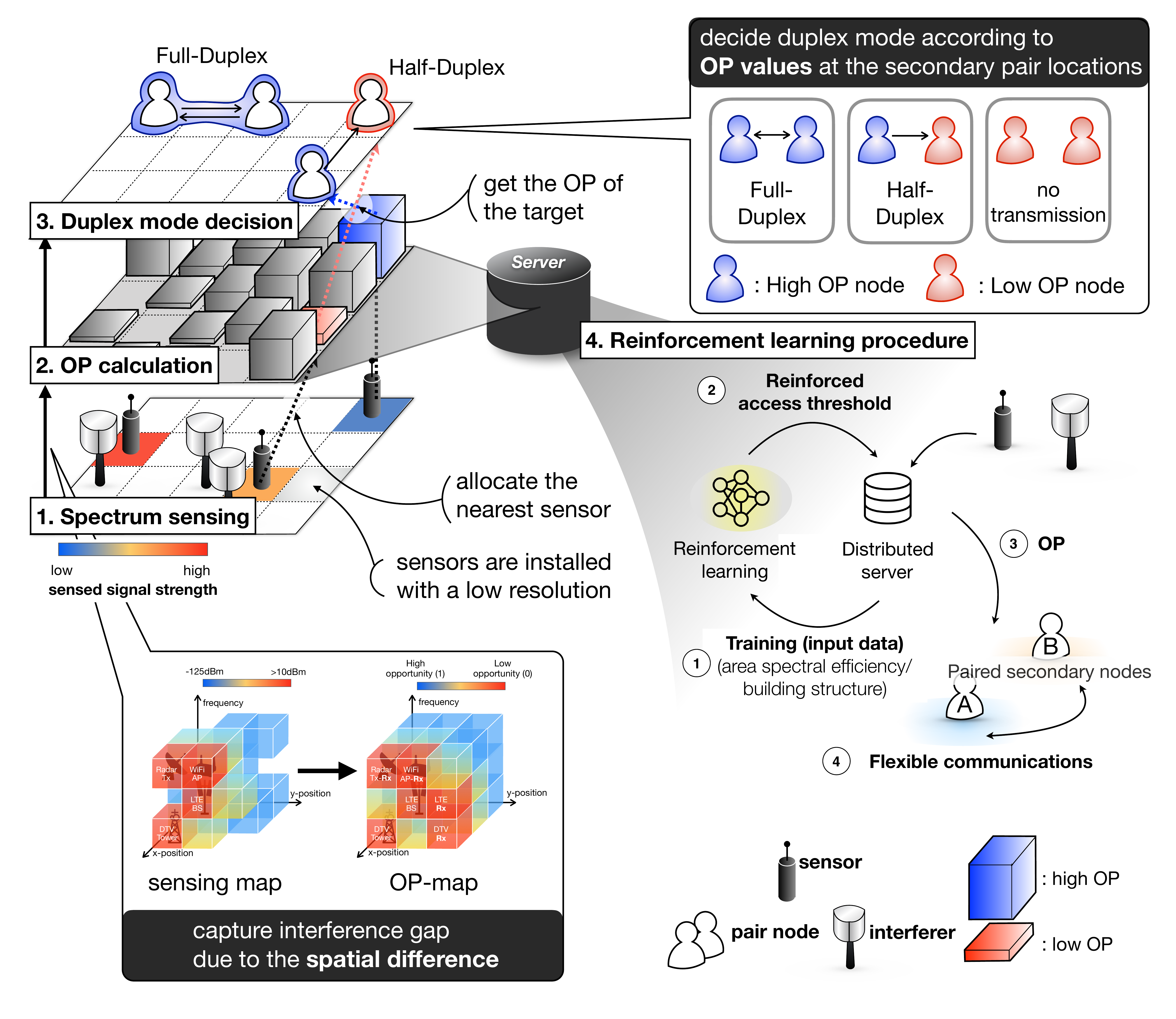}}}
	\caption{A structure of reinforcement learning-based flexible duplex system.}	
	\label{Fig:FD}
\end{figure}

\subsection{System Architecture}

Consider a general spectrum-sharing system composed of communication nodes, spectrum sensors, and a distributed server that only provides the OP value (see Fig.~\ref{Fig:FD}). As illustrated in the figure, the system can be divided into four steps. 
First, deployed spectrum sensors periodically measure the interference level at their own locations and send this to the distributed server. 
Then, the server calculates the OP value to the secondary nodes by the given access threshold from the reinforcement algorithm (The initial access threshold result is gained by pretraining).
Each secondary pair makes a decision based on this OP map and give feedback (spectral efficiency and whether to transmit).
Finally, the server calculates an updated access threshold based on the feedback and sensing information.

%

\subsection{Software/Hardware Layout}

\begin{figure}[t]
	\begin{center}
		\includegraphics[width=3.3in]{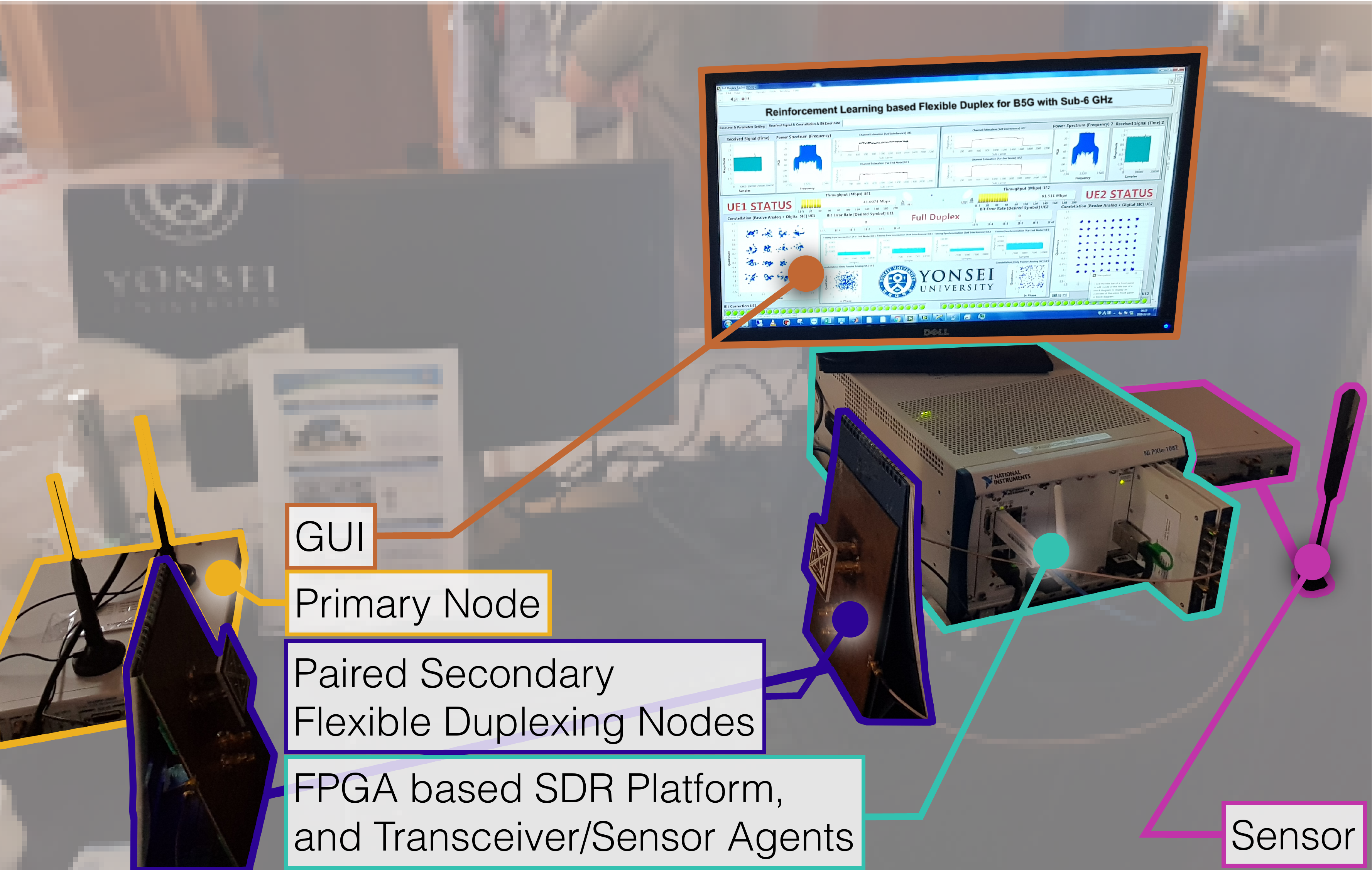}
	\end{center}
	\caption{A real-time demonstration set-up.}\label{Fig:demo}
\end{figure}

\begin{figure}[!t]
	\centering
	\includegraphics[width=1.0\columnwidth,keepaspectratio]
	{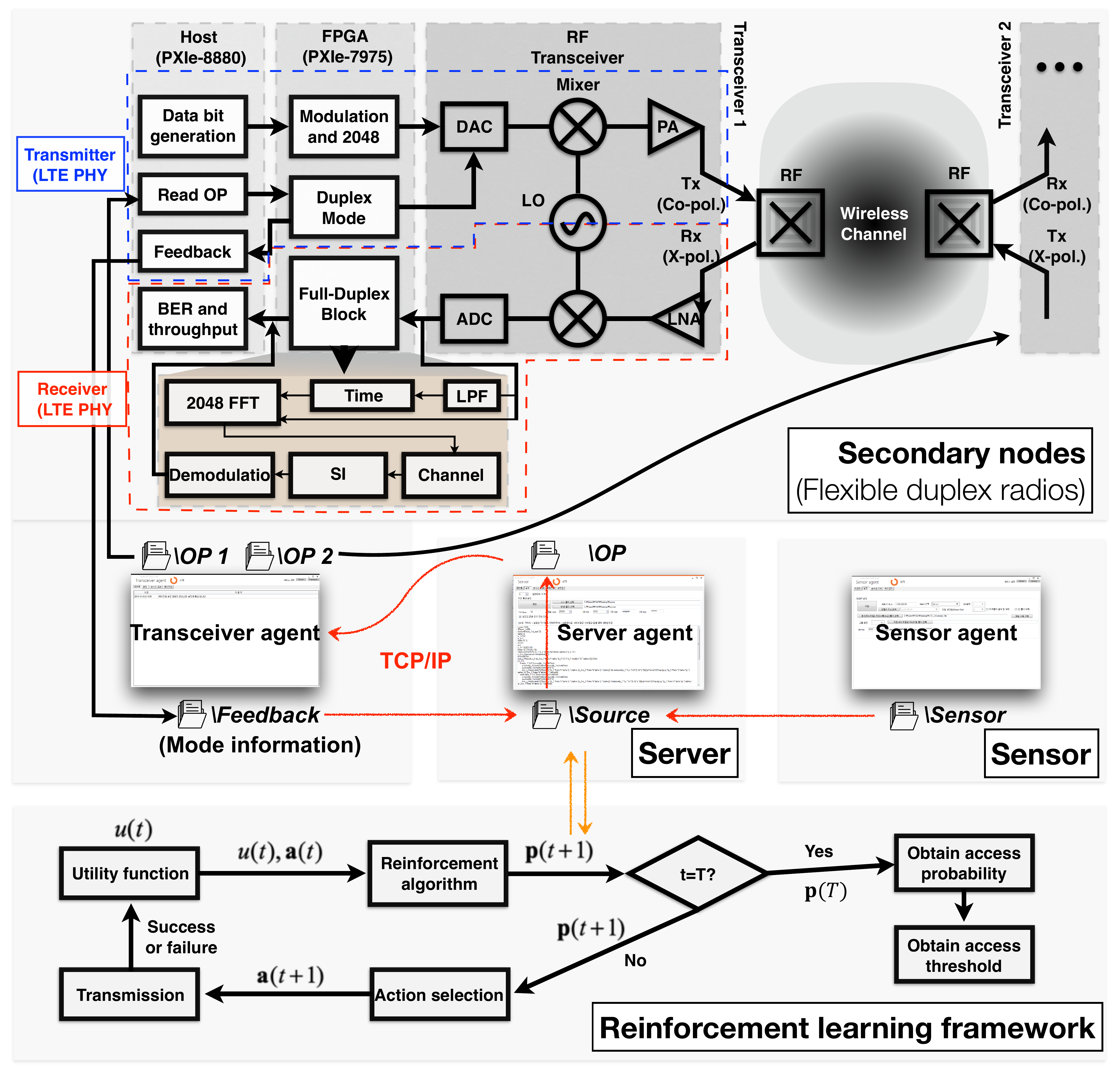}
	\caption{A block diagram with the reinforcement learning framework.}
	\label{Fig:block}
\end{figure}

As Fig.~\ref{Fig:block} shows, our proposed system layout can be divided into two parts: 1) the agent programs with the main server that calculates OP and the access threshold by the reinforcement algorithm; 2) the communication nodes that practically transmit or receive the signal. 
In the main server section, we used Matlab code to design the algorithms for simplicity and low complexity. 
Note that the system latency is limited by the Transmission Control Protocol/Internet Protocol (TCP/IP), and Matlab calculations (They cost about 1~ms and 132~ms, respectively, which dominates over the proposed testbed system, while communication nodes cost about 10~ms for each cycle). This can be further optimized by implementing all these procedures in an FPGA chip and applying the low bound of OP calculation (17~ms).
In the communication nodes section, all nodes are implemented using LabVIEW system design software and the FPGA-based PXIe software-defined radio (SDR) platform as shown in Fig.~\ref{Fig:demo}. 

The flexible duplex platform consists of four main components: a dual-polarized antenna, a controller (PXIe-8880) that operates in the host part, an RF transceiver module (NI 5791), and the FPGA module (PXIe-7975). USRP 2953R and 2922 are utilized as the primary node and sensor, respectively.
In addition, several blocks are implemented in the FPGA domain using Xilinx and LabVIEW IPs (as shown in Fig.~\ref{Fig:block}).
More detailed parameters and hardware specifications are described in~\cite{opmapfd}.
Note that our set-up only requires about $1~m^2$ as shown in Fig.~\ref{Fig:demo}, and a few minutes to install and pretrain the algorithm.

\subsection{Learning and Test Procedures}

In the reinforcement learning section, the secondary transmitters (STXs) learn the optimal spectrum access threshold based on the REINFORCE~\cite{Phansalkar95} algorithm. 
During the learning procedure, every STX attempts transmission by means of slotted ALOHA protocol. With the results of transmission, each STX adjusts its internal state~\cite{Phansalkar95} to maximize the area spectral efficiency. 

Initially, each STX calculates the opportunity~\cite{op} with the aggregate interference value obtained from the nearest spectrum sensor. The initial access probability and internal state are defined by this opportunity value. 
STXs attempt transmission with the initial access probability.
After the transmission, they recorded and shared the information about whether to transmit and the transmission's success or failure. 
With this information, each STX updated its internal state and calculated new access probabilities with an updated internal state. 
With this updated probability, the STXs calculated back the access threshold by setting the opportunity with the learned access probability.


We present the measurement campaign conducted in Veritas Hall Building C 114 at Yonsei University (7.9 m $\times$ 8.6 m). In addition, we measured the system-level evaluations with the building C that was modeled as a real building. 
We obtained the system performance from the BER calculation that considered several \emph{overhead factors}. 
Note that several specifications are set to the LTE standard. 


\section{Evaluations and Conclusions}

This paper proposes the flexible duplex systems that utilizes the reinforcement learning to maximize the area spectral efficiency for the first time. 
First, we thoroughly validated the feasibility of the real-time demonstration that considered both flexible duplexing and the reinforcement learning algorithm.
Through link-level evaluations, our proposed system could generally guarantee improved performance compared to previous research that is the conventional op-map-based flexible duplex system~\cite{dyspandemo} in several conditions.
From extensive system-level evaluations, we compared the suggested algorithm with the result derived from the numerical analysis. Detailed results will be described in further research.
We hope that our research can provide proper insights into prototyping with reinforcement learning as we place our collective shoulder to the wheel of wireless communication systems.

\section{Acknowledgement}
This work was supported by IITP grant funded by MSIT, Korea (No.2018-0-00923, 2018-0-00170), and ICT Consilience Creative Program (IITP-2019-2017-0-01015).


%

\clearpage

\end{document}